%% file: main.tex
\documentclass[aps,prd,twocolumn,groupedaddress,nofootinbib]{revtex4}

\input{config}

\begin{document}

\title{The End of a Black Hole's Evaporation -- Part I}

\author{Fabio D'Ambrosio${}^a$, Marios Christodoulou${}^{b,c}$, Pierre Martin-Dussaud${}^{d,e}$, Carlo Rovelli${}^{e,f,g}$ and Farshid Soltani${}^{h,l}$} 

\affiliation{\vspace{.25cm}${}^a$ Institute for Theoretical Physics, ETH Zurich, Wolfgang-Pauli-Strasse 27, 8093 Zurich, CH}
\affiliation{${}^b$ Department of Computer Science,
	The University of Hong Kong, 
	Pokfulam Road,
	Hong Kong}
\affiliation{${}^c$ Clarendon Laboratory,
	Department of Physics, 
	University of Oxford,
	Oxford,
	United Kingdom}
\affiliation{${}^d$ Institute for Gravitation and the Cosmos, The Pennsylvania State University, University Park, Pennsylvania 16802, USA}
\affiliation{${}^e$ 
Aix Marseille Univ, Universit\'e de Toulon, CNRS, CPT, Marseille, France}
\affiliation{${}^f$ Perimeter Institute, 31 Caroline Street N, Waterloo ON, N2L2Y5, Canada} 
\affiliation{${}^g$ The Rotman Institute of Philosophy, 1151 Richmond St.~N London  N6A5B7, Canada}
\affiliation{${}^h$ Department of Applied Mathematics, University of Western Ontario, London, ON N6A 5B7, Canada}
\affiliation{${}^l$ Dipartimento di Fisica, Universit\`a La Sapienza, I-00185 Roma, EU
\vspace{.15cm}}

\date{\small\today}

\begin{abstract}
\noindent
At the end of the Hawking evaporation the horizon of a black hole enters a physical region where quantum gravity cannot be neglected. The physics of this region has not been much explored. We characterise its physics and introduce a technique to study it. 

\end{abstract}

\maketitle 

\section{Introduction}

\noindent
In a spacetime formed by gravitationally collapsed matter, there are three distinct regions in which curvature becomes Planckian. We expect the approximation defined by quantum field theory interacting with classical general relativity to break down in all three of them. The physics of these regions is quite different.

\begin{figure}
    \centering
    \includegraphics[scale=0.14]{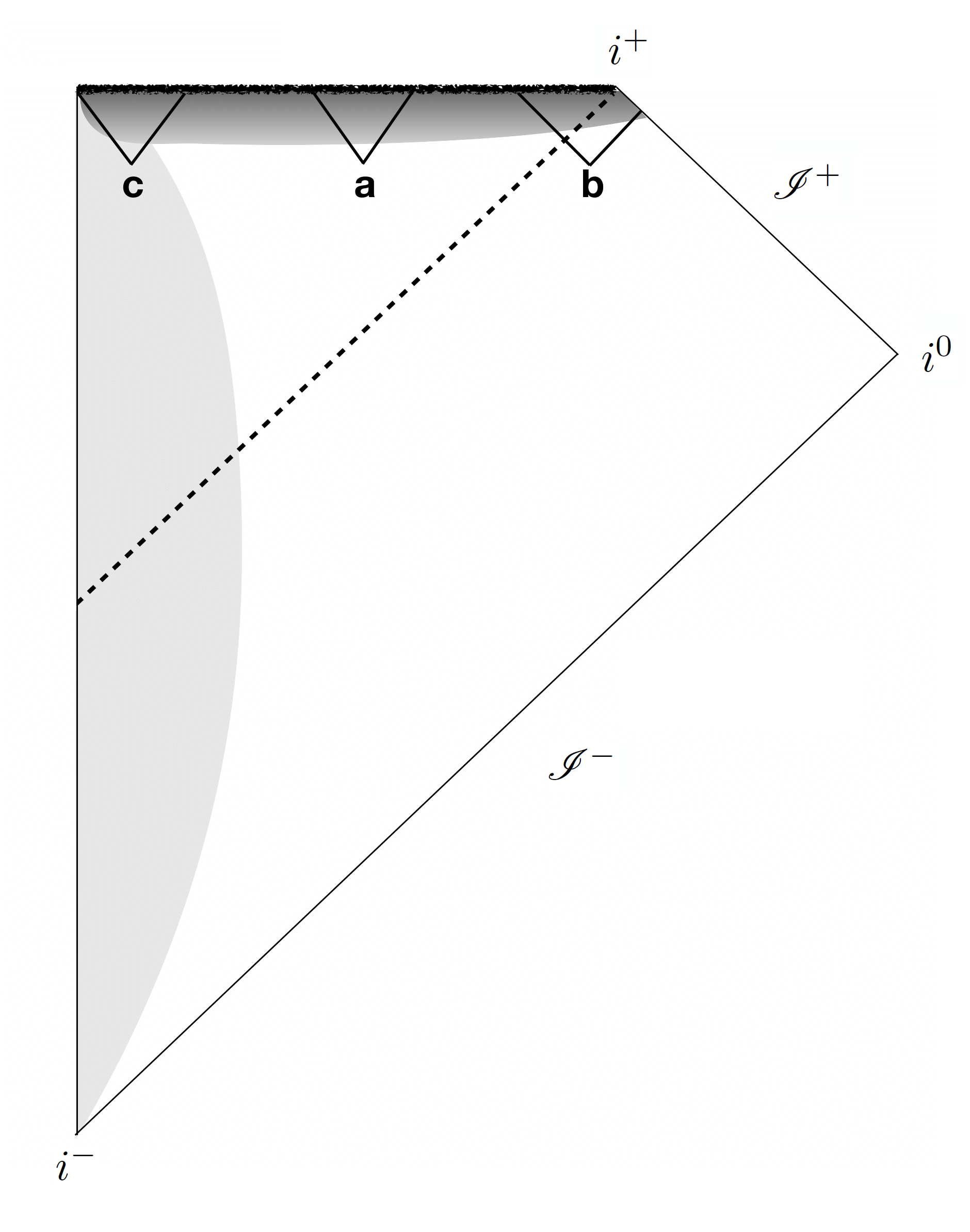}
    \caption{\emph{The three regions of a black hole spacetime where quantum gravity becomes relevant. In the dark grey region quantum gravity cannot be neglected and the diagram itself becomes unreliable. The future of the locations $\mathbf a$, $\mathbf b$ and $\mathbf c$ encounter different quantum gravity phenomena depending, respectively, on the presence of the collapsing matter} ($\mathbf C$), \emph{the horizon} ($\mathbf B$), \emph{or neither} ($\mathbf A$). } 
    \label{fig:ABC}
\end{figure}

The three regions are illustrated in the Carter-Penrose causal diagram of Figure~\ref{fig:ABC}. The dark grey area is the region where quantum gravity cannot be neglected and the diagram itself becomes unreliable. The light grey area is the collapsing matter and the dashed line is the (trapping) horizon (the event horizon is not determined by classical physics). The three physically distinct regions where curvature becomes Planckian are: 
\begin{enumerate}\addtolength{\itemsep}{-1mm}
\item Region $\mathbf C$, in the future of the event $\mathbf c$ in the diagram, which is directly affected by the collapsing matter reaching Planckian density.
\item Region $\mathbf B$, in the future of the event $\mathbf b$ in the diagram, which is affected by the horizon reaching Planckian size because of Hawking's evaporation.
\item Region $\mathbf A$, in the future of any location like $\mathbf a$ (that is a generic event in the dark grey area distant from the events $\mathbf b$ and $\mathbf c$) in the diagram, where the curvature becomes Planckian but the classical evolution to the singularity is not causally connected to the collapsing matter or to the horizon. 
\end{enumerate}

The physical distance between these regions depends on the age of the black hole at the time when its horizon reaches the quantum region. This age depends in turn on the overall mass of the black hole \emph{before} being shrunk by Hawking evaporation.

To give a rough estimate of these distances we consider for simplicity the interior of a Schwarzschild black hole. (Most of the evaporation takes place at late times.) The line element is
\begin{multline}
    \dd s^2 = - \left(1 - \frac{2 G m}{r} \right) \dd t^2 + \left( 1 - \frac{2 G m}{r} \right)^{-1} \dd r^2 \\ + r^2 \left( \dd \theta^2 + \sin^2\theta\, \dd \phi^2 \right)
\end{multline}
We can take the three locations $\mathbf a$, $\mathbf b$ and $\mathbf c$ to be at the same fixed values of $\theta, \phi, r$ and at three different values $t_{\mathbf a}, t_{\mathbf b}, t_{\mathbf c}$ of the $t$ coordinate. The proper distance $\dd l$ along a line of constant $\theta,\phi,r$, namely a nearly horizontal line in the causal diagram, is given by the line element
\begin{equation}
    \dd l = \dd t \sqrt{\frac{2Gm}{r} - 1}\  \, .
\end{equation}
The quantity $\dd l$ becomes large approaching the quantum gravitational dark grey region of Figure~\ref{fig:ABC}. Curvature scalars behave as $\sim m/r^3$ and hence become Planckian at $r/L_{Pl}\sim (m/M_{Pl})^{1/3}$ where $L_{Pl}$ and $M_{Pl}$ are the Planck length and the Plank mass, giving
\begin{eqnarray}
\dd l\sim\sqrt2\ (m/M_{Pl})^{1/3} \dd t
\end{eqnarray}
near the quantum gravitational dark grey region of Figure~\ref{fig:ABC}. 
For a stellar mass ($m\sim M_{\odot} \sim 10^{38} M_{Pl}$) black hole, if no further mass enters the horizon, the end of the Hawking evaporation is at $t_{\mathbf b}-t_{\mathbf c}\sim (M_\odot/M_{Pl})^3 L_{Pl}$, hence the distance between $\mathbf b$ and $\mathbf c$ is
\begin{eqnarray}
L\ \sim \ L_{Pl}{\left(\frac{M_{\odot}}{M_{Pl}}\right)^{\frac{10}{3}}}\sim \ 10^{75} \ {\rm light\ years},
\end{eqnarray}
which is huge. That is: the locations $\mathbf b$ and $\mathbf c$ are extremely distant from each other~\cite{Perez2015c,Christodoulou2015,Bengtsson2015,Christodoulou2016a}.  This is a rough estimate, but the conclusion is general: the distance between $\mathbf b$ and $\mathbf c$, which is to say the "depth" of the black hole, is huge, for an old black hole. Notice that what makes this distance large is not the smallness of the $r$ coordinate considered (which is not Planckian): rather, it is the long lifetime of the black hole that builds up the length.

It is worthwhile pausing to ponder this fact: near the end of the Hawking evaporation of an isolated stellar-size black hole, the collapsing matter entering the quantum region is at a ---\emph{spatial, not temporal!---} distance of $10^{75}$ light years from the horizon. Locality demands the physics of spatially widely separated phenomena to be independent. It is reasonable to expect quantum gravity to affect the causal structure of spacetime, but in small fluctuations, not by suddenly causally connecting events that are extremely far apart. 

It follows that the physics of each of the three regions $\mathbf A$, $\mathbf B$ and $\mathbf C$ can be studied independently from the others (until something brings them in causal contact). More precisely, the physics of the $\mathbf A$ region can be studied independently from what happens at $\mathbf B$ or $\mathbf C$; while these depend on the physics of the $\mathbf A$ region, since this bounds the horizon and the collapsing matter. 


\begin{figure}
    \centering
    \includegraphics[scale=0.14]{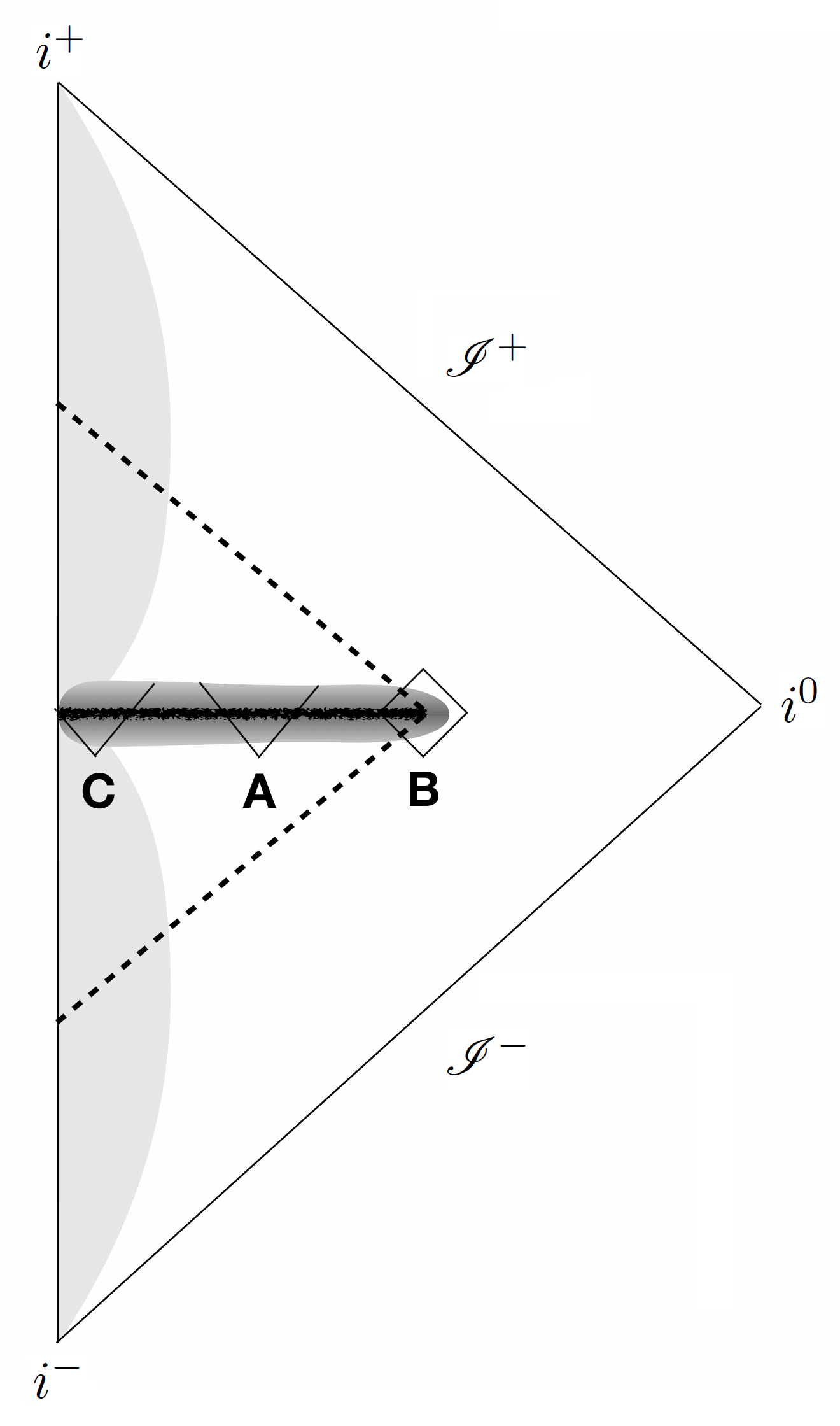}
    \caption{\em Carter-Penrose causal diagram of the black to white transition.} 
    \label{fig:fst}
\end{figure}

Since the physics of the $\mathbf A$ region neither depends on the collapsing star nor on the shrinking horizon, it can be studied in the context of an eternal black hole. This setting gets rid of the collapsing matter, and allows is to neglect the Hawking radiation, whose back-reaction shrinks the area of the horizon until it enters the quantum region. There is an extensive recent literature on the possible scenarios for the physics of the $\mathbf A$ region. A much studied possibility is that spacetime continues on the future of the would-be singularity, namely on the future of the dark grey region of Figure~\ref{fig:ABC}, into an anti-trapped region, namely a region with the metric of a white hole~\cite{Modesto2004,Ashtekar2005a,Modesto2006,Modesto2006b,Gambini:2008bh,Corichi2015,Ashtekar2018d,
Clements2019,Bodendorfer2019,Ashtekar2020,Gambini2020}. 

Here we take this possibility as an assumption. This seems by far the most plausible scenario, the one which is more coherent with the physics that we know. In fact, the entire effect of quantum gravity is a slight violation of the Einstein equations in the high curvature region, which prevents curvature to diverge and allows spacetime to continue. This possibility was noticed long ago, already in the fifties, by Synge~\cite{Synge1950}.

Following in particular~\cite{Haggard2014,DeLorenzo2016,DAmbrosio2018a,Bianchi2018c}, we assume here that the anti-trapped region is bounded by a future horizon that connects it to the region external to the black hole, a surprising possibility first noticed in~\cite{Haggard2014}. The full spacetime has therefore the causal structure depicted in the Carter-Penrose diagram in Figure~\ref{fig:fst}. 
   
The $\mathbf C$ region is where the collapsing matter itself reaches the quantum gravity regime. It is called the "Planck star" phase of the collapsing matter~\cite{Rovelli2014}. Here, following~\cite{Rovelli2014}, we simply assume that some form of matter bounce compatible with this scenario happens.
   
In this paper we focus on the physics of the $\mathbf B$ region, namely on the events near the end of the Hawking evaporation of the black hole. This is the region where the trapping horizon tunnels into an anti-trapping horizon. Covariant Loop Quantum Gravity (LQG)~\cite{Rovelli2015} can be utilised to study the region around the classical singularity using the spinfoam formalism~\cite{Perez:2012wv}. The transition amplitude for the entire quantum region (the whole dark grey region in Figure~\ref{fig:ABC}) was first roughly estimated using LQG in~\cite{Christodoulou2016,Christodoulou2018}. Here we use a similar technique to begin a more refined study of the $\mathbf B$ region only.

In particular, in section~\ref{section:II} we compute the classical intrinsic and extrinsic geometry of a boundary of the $\mathbf B$ region in terms of a small number of parameters characterising the spacetime and the transition. The quantum transition amplitude that describes the $\mathbf B$ region is going to be a function of these parameters. Furthermore, in view of the spinfoam transition amplitude calculations, in section~\ref{section:dual_triangulation} we introduce and study a triangulation of the boundary of $\mathbf B$ and its discrete geometry. In a forthcoming companion paper we introduce a full discretisation of the $\mathbf B$ region compatible with the triangulation of its boundary introduced in section~\ref{section:dual_triangulation} and we use it to explicitly write the transition amplitude for the phenomenon in terms of LQG (spinfoam) techniques.

The main result of this paper is the identification of the four parameters which characterise the quantum transition at the $\mathbf B$ region and the definition of the corresponding transition amplitude as a function of these parameters. The actual computation of this amplitude will be addressed in the forthcoming companion paper.

\section{The boundary of the \texorpdfstring{$\mathbf B$}{\textbf{B}} region}
\label{section:II}

\noindent
To study the $\mathbf B$ region, we restrict for simplicity to the spherically symmetric case and we assume the rest of spacetime to be classical. For the most part, this is a good approximation, since the curvature is below Planckian values and quantum effects are likely to be negligible. This is however not true for the boundary between the $\mathbf B$ and the $\mathbf A$ region. We therefore simplify the problem by describing the $\mathbf A$ region with an effective classical metric, as in~\cite{DAmbrosio2018a}. 

An effective metric for the entire spacetime that takes into account the effect of Hawking radiation was studied in~\cite{martin-dussaud2019a}. Here we further simplify this scenario by disregarding the presence of Hawking radiation in the last phases of the evaporation, hence around the $\mathbf B$ region, in spite of the radiation being strong in this region. We do not know how good this approximation is. We assume that the black hole has already evaporated to a small size and we take the metric around the $\mathbf B$ region to be well approximated by a Schwarzschild metric, up to quantum corrections in the $\mathbf A$ region.

The effective geometry of the $\mathbf A$ region continuing from the trapped to the anti-trapped region can be described by the line element~\cite{DAmbrosio2018a}
\be
\dd s_l^2= - \frac{4(\tau^2+l_{})^2}{2m-\tau^2}\dd\tau^2 + \frac{2m-\tau^2}{\tau^2+l_{}}\dd x^2 + 
(\tau^2+l_{})^2\dd\Omega^2 \label{me2}\, ,
\ee 
where $\dd\Omega^2$ is the metric of the 2-sphere, $l\ll m$ is an intrinsic parameter of the effective metric and $-\sqrt{2m} < \tau < \sqrt{2m}$. This line element defines a genuine pseudo-Riemannian space, with no divergences and no singularities $\forall \,l \neq 0$.

In the limit $l\to0$, the metric locally converges to the interior Schwarzschild metric for a black hole in $-\sqrt{2m} < \tau < 0$ and to the interior Schwarzschild metric for a white hole in $0< \tau < \sqrt{2m}$. This can be easily seen by performing the following change of coordinates:
\be
r=\tau^2  \quad \text{and} \quad t=x\, ,
\ee
where $r$ and $t$ are the usual Schwarzschild coordinates. In this limit, $\tau=0$ becomes the singularity separating the trapped from the anti-trapped region. For $l\ne 0$ the curvature remains instead bounded. 

\begin{figure}
\includegraphics[height=2.3cm]{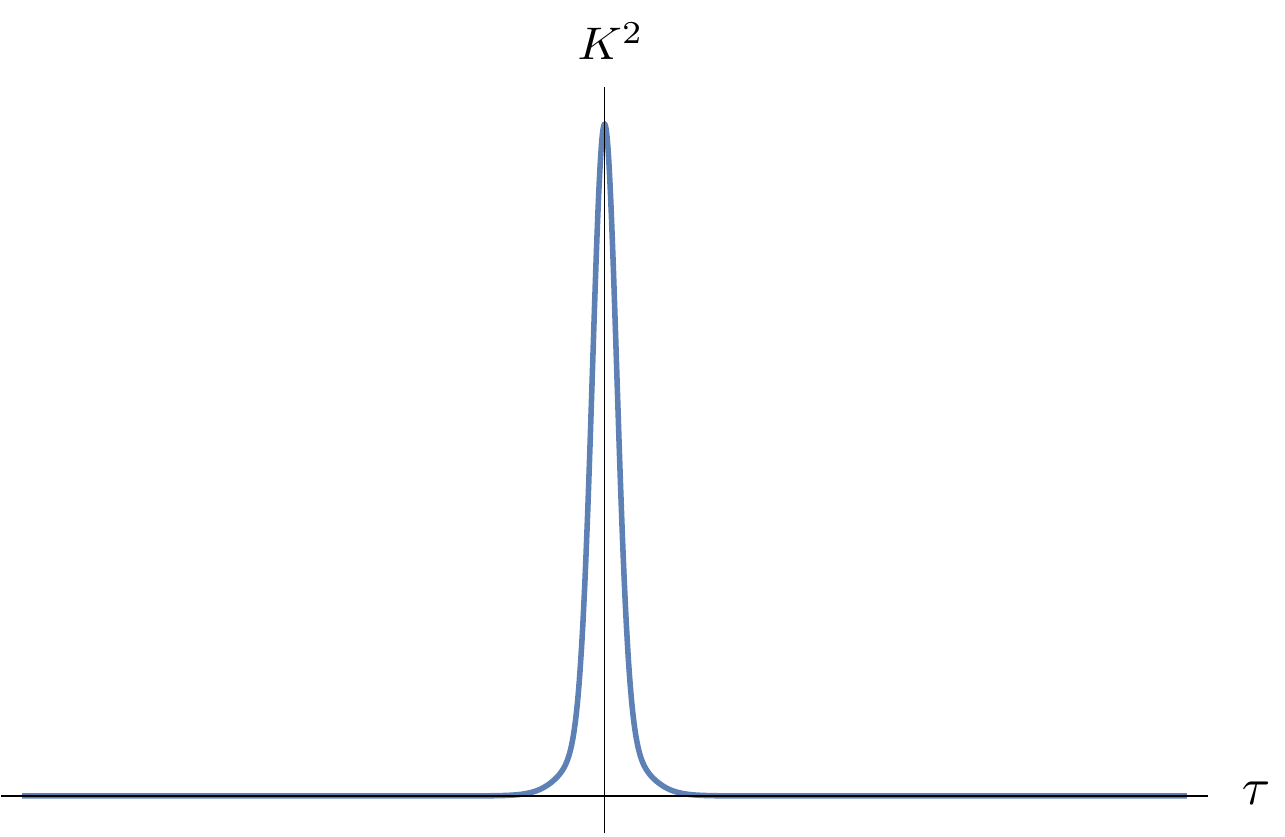}
\vspace{-1em}
\caption{\em The bounded curvature scalar~\eqref{boundedDive}.}
\label{fig:K_curvature_scalar}
\end{figure}

Up to terms of order $\mathcal{O}\left({l_{}}/{m}\right)$, the curvature scalar $K^2\sim R_{\mu\nu\rho\sigma}R^{\mu\nu\rho\sigma}$, which is plotted in Figure~\ref{fig:K_curvature_scalar}, is
\begin{eqnarray}
K^2(\tau)&=& \frac{9\,  l_{}^2 - 24 \,  l_{} \tau^2+ 48\, \tau^4}{  (l_{} + \tau^2)^8}m^2 \,.
\label{boundedDive}
\end{eqnarray}
It has \emph{finite} maximum value 
\be
K^2(0)=\frac{9\, m^2}{ l_{}^6} .
\ee
The Ricci tensor vanishes up to terms of order~$\mathcal{O}(l_{}/m)$.

The space-like surfaces $\tau = \text{constant}$ can be used to foliate the interior of both the black and the white hole. Each of these surfaces has the topology $S^2 \times \mathbb{R}$. Suppressing one angular coordinate, they can be depicted as long cylinders of different radii and heights. In the interior black hole region ($-\sqrt{2m} < \tau < 0$), as $\tau$ increases, the radial size of the cylinder shrinks while the axis of the cylinder gets stretched. At $\tau = 0$ the cylinder reaches a minimal width and maximal length, and then smoothly bounces back and starts increasing its radial size and shrinking its length as $\tau$ increases in the interior white hole region ($0 < \tau < \sqrt{2m}$). The cylinder inside the hole never reaches arbitrary small sizes (the singularity), but it rather bounces at a small finite radius $l_{}$.

The value of $l_{}$ can be roughly fixed by the requirement that $K(0) \sim 1$ (in Planck units), which gives $l \sim m^{1/3}$. This means that the bounce happens at a larger scale than the Planck one. The limit $l_{}\to 0 $ is simply the joining of a Schwarzschild black hole interior and a Schwarzschild white hole interior through the singularity. This is not a Riemann space --- it is analogous to a double cone: a space with a singular region of measure zero --- but it is a rather well behaved metric manifold, where geodesics can be defined and studied~\cite{DAmbrosio2018a}. 

The presence of a minimum finite radius $l_{}$ in the $\mathbf A$ region has far-reaching consequences for the physics of the $\mathbf B$ region. 

\subsection{Choice of the boundary}

\noindent
The idea to define a boundary for the $\mathbf B$ region is to first surround it in the causal diagram with a diamond shaped null surface $\Sigma$ (see Figure~\ref{fig:fst}), that is a diamond null surface times a sphere in spacetime, and then, since an appropriate boundary for computing transition amplitudes must be spacelike, to slightly deform $\Sigma$ into a spacelike surface. This surface is the Heisenberg cut we choose, namely the surface we shall take as the boundary between the quantum and the classical regions. Notice that in quantum gravity, the Heisenberg cut is also a spacetime boundary (see~\cite{Rovelli:2004fk}, section 5.6.4). 

We want now to concretely specify $\Sigma$ and compute its intrinsic and extrinsic geometry. Since it has been assumed that the dissipative irreversible physics of the Hawking radiation is over at this point, the $\mathbf B$ region must be time-reversal invariant. The surface $\Sigma$ can consequently be seen as the union of two surfaces, a \emph{past} one $\Sigma^p$ and a \emph{future} one $\Sigma^f$, equal up to time reflection, $\Sigma=\Sigma^p\cup\Sigma^f$. Here, the labels $p$ and $f$ stand for past and future, and later on we shall also use the index $t= \{ p,f \}$ (hence $\Sigma^t$) where $t$ stand for \emph{time}. Accordingly, we only need to study the past boundary $\Sigma^p$, as the future boundary $\Sigma^f$ is determined by symmetry.   

The metric around the $\mathbf B$ region is assumed to be well approximated by the Schwarzschild metric up to quantum corrections in the $\mathbf A$ region. The past boundary is then contained in the external and in the black hole regions of a Kruskal diagram representing Schwarzschild spacetime. Since both regions are covered by the ingoing Eddington-Finkelstein coordinates, we can use these coordinates to define $\Sigma^p$. The line element in these coordinates reads
\be
\dd s^2=-\left(1-\frac{2m}{r}\right)\dd v^2+2\dd r\,\dd v+r^2\,\dd\Omega^2.
\label{equation:metric}
\ee 
The Schwarzschild time coordinate $t$ is related to the ingoing Eddington–Finkelstein coordinates by 
\be
t=v-r^*=v-r-2m\ln\left|\frac{r}{2m}-1\right|\,,
\ee 
or
\be
\dd t=\dd v-\frac{\dd r}{\left(1-\frac{2m}{r}\right)}.
\ee 

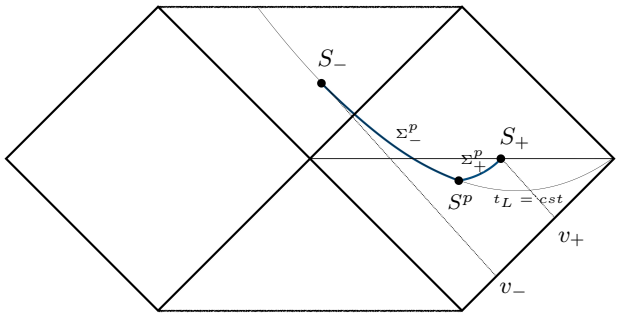
\begin{figure}
\centering
\begin{overpic}[width=1 \columnwidth]{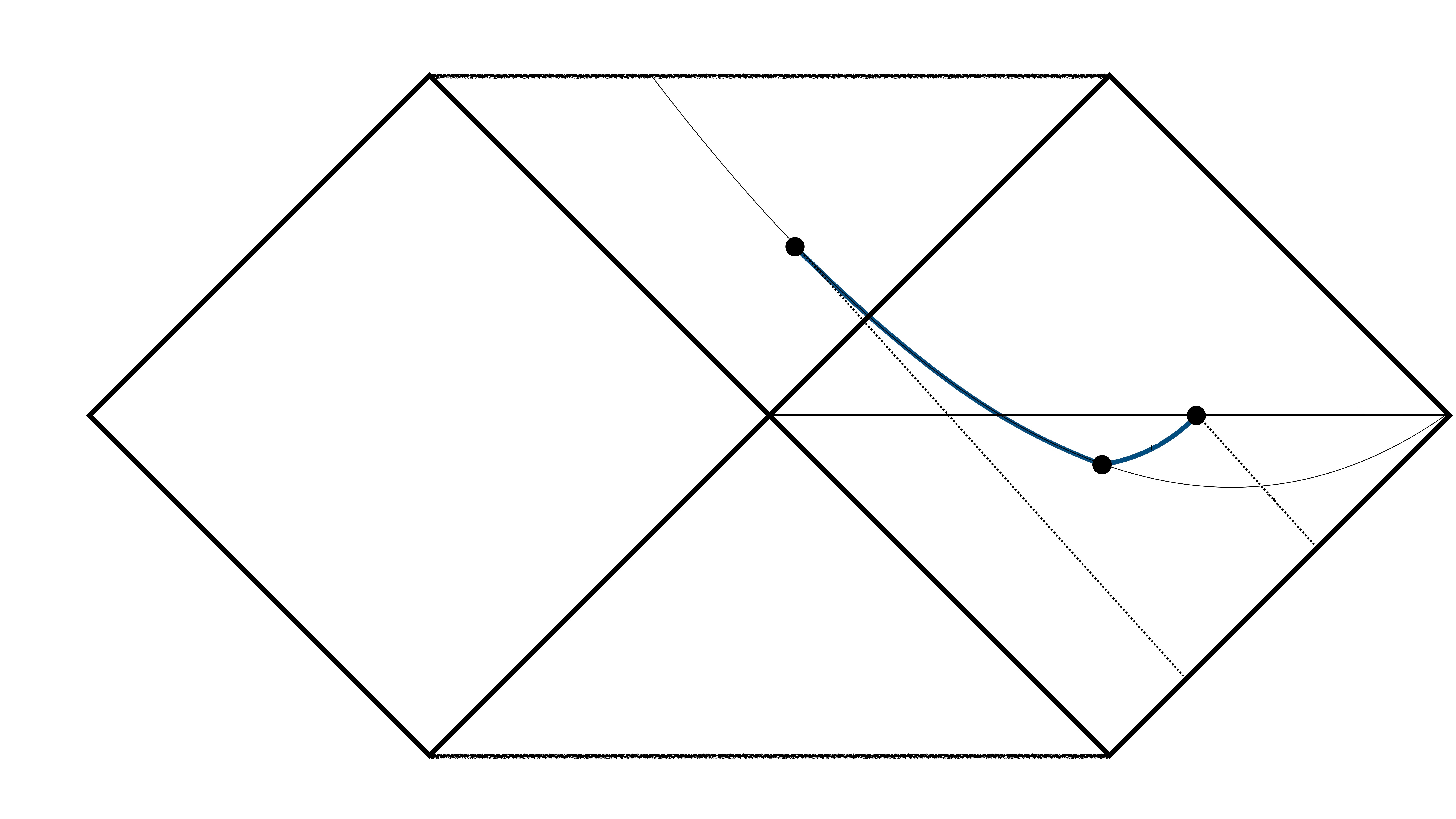}
\put (74,20) {\footnotesize $S^p$}
\put (82,30) {$S_+$}
\put (54,42) {$S_-$}
\put (66,31) {\tiny $\Sigma_-^p$}
\put (76,27) {\tiny $\Sigma_+^p$}
\put (91,15) {$v_+$}
\put (82,7) {$v_-$}
\put (81,21) {\tiny $t_L = cst$}
\end{overpic}
\caption{The past portion of the boundary surface.}
\label{fig:boundary_in_kruskal_diagram}
\end{figure}

The null past diamond boundary can be defined in the Kruskal diagram as follows. Let $S_{out}$ be a point (a two-sphere in spacetime) outside the horizon at advanced time $v_+$ and Schwarzschild time $t=0$. Let $S_{in}$ be a point inside the horizon at advanced time $v_-<v_+$ and Schwarzschild radius $r_-= l$. The null past diamond boundary is taken to be the union of the outgoing past light cone of $S_{out}$ and of the ingoing past light cone of $S_{in}$ from their intersection upward; see Figure~\ref{fig:boundary_in_kruskal_diagram}. 

Note that the presence of a minimum finite radius $l_{}$ in the $\mathbf A$ region fixes the value $r_-$ of the Schwarzschild radius of the point $S_{in}$. Furthermore, being the Schwarzschild radius $r$ a temporal coordinate inside the black hole, while the radius $r_+$ of $S_{out}$ is a measure of the spatial coordinate distance of $S_{out}$ from the horizon, the radius $r_-= l$ must not be interpreted as a measure of the spatial coordinate distance of $S_{in}$ from the horizon but as the minimal internal radius reached by the black hole in region $\mathbf A$.

To simplify the notation, in the following we replace the labels \emph{out} and \emph{in} with the index $\pm= \{+,-\} \equiv \{{out},{in}\}$, e.g. $S_+\equiv S_{out}$ and $S_-\equiv S_{in}$.  

Next, we define the spacelike past boundary $\Sigma^p$ by slightly deforming the null past diamond boundary while keeping fixed $S_+$ and $S_-$. A convenient choice of deformation is the following one. Consider the surface $\Sigma^p_-$ of constant Lema\^{i}tre time coordinate \cite{Lemaitre1933,Blau:fk}
\be
t_L=t+2\sqrt{2mr}+2m\ln\left|\frac{\sqrt{r/2m}-1}{\sqrt{r/2m}+1}\right|, 
\label{equation:lemaitretime}
\ee
passing by $S_-$ and the surface $\Sigma^p_+$ defined by 
\be
v - \beta r = const, 
\label{equation:vbetar}
\ee passing by $S_+$, for some constant $\beta \in \mathbb{R}$. Let $S^p$ be their intersection; see Figure~\ref{fig:boundary_in_kruskal_diagram}. We choose the spacelike past boundary $\Sigma^p$ to be the union of the portion of $\Sigma^p_-$ between $S^p$ and $S_-$ and the portion of $\Sigma_+$ between $S^p$ and $S_+$. The parameter $\beta$ can be fixed by requiring the continuity of the normal to $\Sigma^p$ at $S^p$.

\begin{figure}[h]
    \centering
    \includegraphics[scale=0.13]{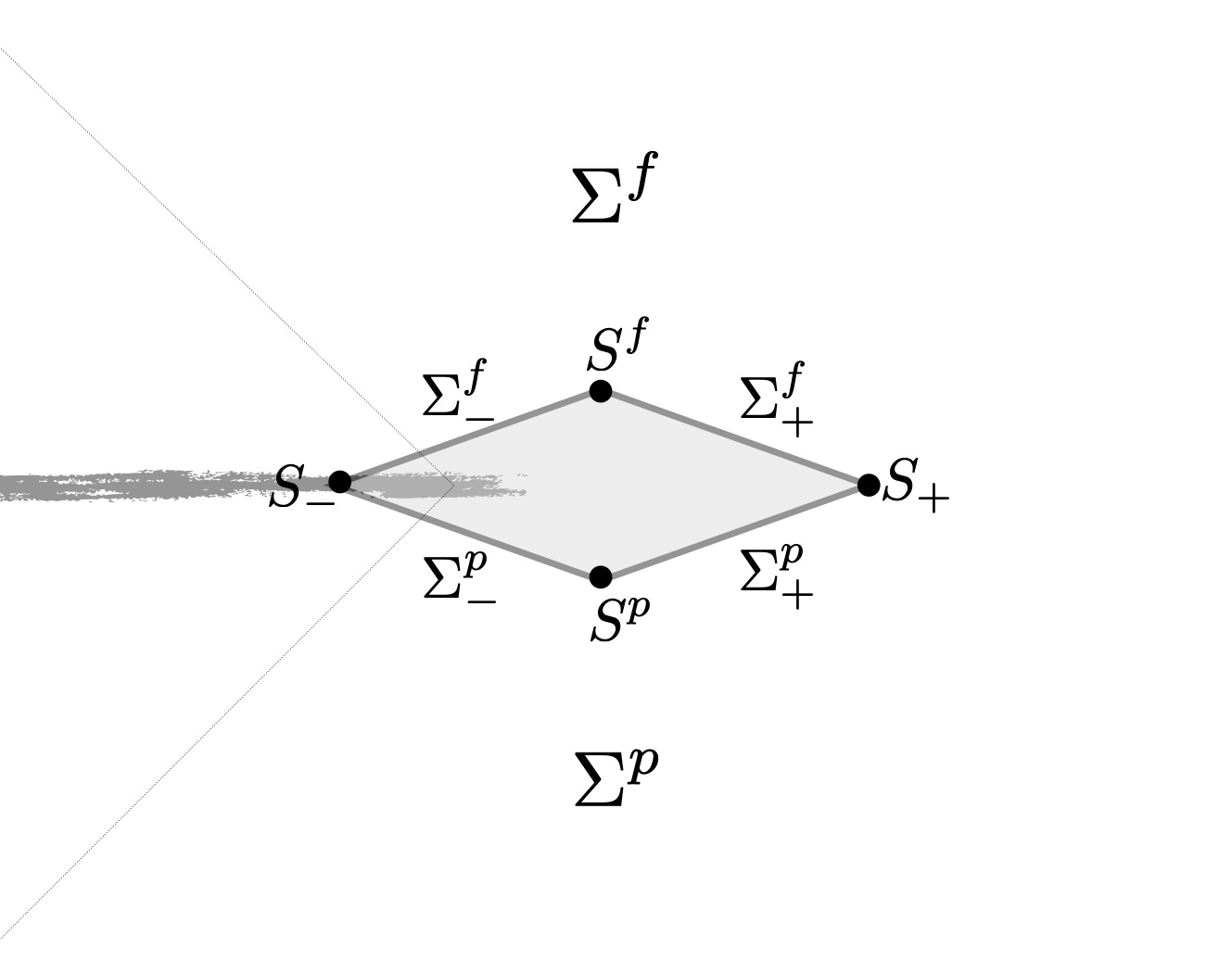}
    \caption{Carter-Penrose diagram of the $\mathbf B$ region with the surface $\Sigma$ and its components highlighted.}
    \label{fig:diamond}
\end{figure}

The spacelike future boundary surface $\Sigma^f$ is defined to be the time-reversal of the surface $\Sigma^p$ and the full spacelike boundary surface $\Sigma$ is then partitioned in the four components $\Sigma^p_+$, $\Sigma^p_-$, $\Sigma^f_+$ and $\Sigma^f_-$; see Figure~\ref{fig:diamond}. The Carter-Penrose diagram of the $\mathbf B$ region consists of two separate portions of the Kruskal diagram which are appropriately joined. This is the "cutting and pasting" used in~\cite{Haggard2014} in order to write for the first time a metric for the black-to-white transition. 

We now need to determine the intrinsic and the extrinsic geometry of $\Sigma$. 

\subsection{Intrinsic geometry}


\noindent
The intrinsic geometry of $\Sigma^p_+$ is obtained by differentiating its defining equation (equation~\eqref{equation:vbetar}),
\be
\dd v=\beta \dd r,
\ee
and inserting the result in the line element in equation~\eqref{equation:metric}. This gives
\be
\dd s^2_+=\beta\left(2-\beta\left(1-\frac{2m}{r}\right)\right)\dd r^2+r^2\,\dd\Omega^2.
\label{equation:metric+}
\ee

To find the intrinsic geometry of $\Sigma^p_-$, we rewrite the explicit expression of the Lema\^{i}tre time coordinate in equation~\eqref{equation:lemaitretime} in terms of the $(v,r)$ coordinates. Then we differentiate it, finding that on a constant $t_l$ surface the following relation is satisfied:
\be
\dd v=\frac{\dd r}{1+\sqrt{2m/r}}\, .
\ee
Using this relation in the line element in equation~\eqref{equation:metric}, we obtain that the line element resulting from the intrinsic metric of the $\Sigma^p_-$ surface is
\be
\dd s^2_-=\dd r^2+r^2\,\dd\Omega^2.
\label{equation:metric-}
\ee 
That is, $\Sigma^p_-$ is intrinsically flat.

\subsection{Extrinsic geometry}

\noindent
Next, we want to determine the extrinsic geometry of $\Sigma$.

Since the two surfaces $\Sigma^p_+$ and $\Sigma^p_-$ are both defined by constraint equations of the form $C=0$, it is easy to compute their normal 1-forms using
\be
n_\mu = - \frac{\partial_\mu C }{|\partial^\nu C \partial_\nu C|^{1/2}} .
\ee
In Schwarzschild coordinates, the normals to the surfaces $\Sigma^p_-$ and $\Sigma^p_+$ are then given by
\be
 n_\mu^-= \left(-1,-\frac{\sqrt{2mr}}{r-2m},0,0 \right)\, ,
\ee
\be
  n_\mu^+= \frac{\left(-1, \beta - \left(1-\frac{2m}{r}\right)^{-1},0,0 \right)  }{\left|\beta \left( \beta - 2 - \frac{2m\beta}{r} \right)\right|^{1/2}}\,.
\ee
Demanding that the normals match on $S^p$, uniquely fixes the value of $\beta$:
\be \label{beta rS}
\beta = \frac{1}{1+\sqrt{\frac{2m}{r_{S^p}}}}.
\ee


To deal with the extrinsic curvature of the surfaces $\Sigma^p_\pm$ it is easier to express them as systems of parametric equations $x_{\pm}^\mu = x_{\pm}^\mu (y_{\pm}^a)$, where $y_{\pm}^a$ are some parameters which serves as intrinsic coordinates to the surfaces. Given a generic surface defined by the system of parametric equations $x^\mu = x^\mu (y^a)$ for some $y^a$, the tangent 1-form to the surface $e^\mu_a$ is given by
\be
    e^\mu_a = \frac{\partial x^\mu}{\partial y^a}
\ee
and the extrinsic curvature tensor $k_{ab}$ of the surface reads
\be
    k_{ab} = e^\mu_a e^\nu_b \nabla_\mu n_\nu\, .
\ee
Let $k^{\pm}_{ab}$ be the extrinsic curvature of $\Sigma^p_{\pm}$. Then, a straightforward calculation gives
\be
    k^-\equiv k^-_{ab}\, \dd x^a \dd x^b = \sqrt{\frac{m}{2r^3}}\, \dd r^2 - \sqrt{2mr}\, \dd\Omega^2
\ee
and
\be
\begin{split}
    k^+\equiv k^+_{ab}\, \dd x^a \dd x^b = &  \frac{m\beta^{3/2} ( r(3-\beta) + 2m\beta)}{\sqrt{r^5(r(2-\beta)+2m\beta)}}\, \dd r^2 \\
    & - \frac{r(1-\beta)+2m\beta}{\sqrt{\beta(2-(1-2m/r)\beta)}}\, \dd\Omega^2 \, .
    \end{split}
\ee

This completes the computation of the geometry of the boundary of $\mathbf B$. This geometry is entirely determined by four parameters: the mass $m$, the Schwarzschild radii $r_\pm$ of the spheres $S_\pm$, which by construction satisfy
\be
r_-<2m<r_+\, ,
\ee
and the retarded time $v=v_+-v-$. The physical interpretation of these four parameters is transparent. The mass $m$ is the mass of the black hole when the black-to-white transition happens; the retarded time $v$ is the external (asymptotic) time it takes for the transition to happen; the radius $r_+$ is the minimal external radius where we assume the classical approximation to hold; the radius $r_-$ is the minimal internal radius reached by the black hole interior in region $\mathbf A$. When $m$ and $r_\pm$ are fixed, the value of $v$ can be equivalently determined by fixing $\beta$ or $r_{S^p}$. These are the only parameters describing the quantum transition.

Quantum gravity should determine a transition amplitude $W$ for the process as a function of these four parameters
\be
W=W(m,r_\pm,v). 
\ee
In Planck units, the four parameters can be seen as dimensionless. We expect the specific details of the chosen $\Sigma$ not to matter, as they can be absorbed in a shift of the Heisenberg cut (as long as it does not enter the quantum region). 

The task of the forthcoming companion paper is to write an explicit expression for the function $W(m,r_\pm,v)$ using the covariant LQG transition amplitudes. These are given in an expansion in number of degrees of freedom. At finite order, the amplitudes are defined for specific discretisations of the geometry. Below we define a first order discretisation of $\Sigma$ in the form of a triangulation. As we shall see in the companion paper, this triangulation can in fact be seen as the boundary of a cellular decomposition of the $\mathbf B$ region.

\section{The triangulation of \texorpdfstring{$\Sigma$}{\textSigma}}
\label{section:dual_triangulation}

\noindent
The topology of the $\mathbf B$ region is $S^2\times [0,1]\times[0,1]$ and the topology of its boundary $\partial B=\Sigma=\Sigma^p\cup\Sigma^f$ is $S^2\times S^1$. 

We can identify two symmetries of the geometry of $\Sigma$ and one symmetry of its topology:
\begin{itemize} 

\item The $Z_2$ time reversal symmetry that exchanges $p$ and $f$.

\item The $\mathrm{SO}(3)$ symmetry inherited by the spherical symmetry of the overall geometry.

\item A $Z_2$ symmetry that exchanges the internal (minimal radius) sphere $S_-$ and the external (maximal radius) sphere $S_+$. This is a symmetry of the topology, but not of the geometry, since $S_-$ and $S_+$ have different size.

\end{itemize}

To find a triangulation of $\Sigma$ we discretise the two spheres $S_-$ and $S_+$ into regular tetrahedra. This replaces the continuous $\mathrm{SO}(3)$ symmetry with the discrete symmetries of a tetrahedron. In particular, we discretise each of the two spheres $S_\pm$ in terms of a tetrahedron $t_\pm$. We label the four vertices of each tetrahedron as $v_{\pm a}$ where $a=1,2,3,4$; and the triangles bounding the tetrahedra as $\ell_{\pm a}$, where the triangle $\ell_{\pm a}$ is opposite to the vertex~$v_{\pm a}$. 

\begin{figure}
    \centering
    \includegraphics[scale=0.12]{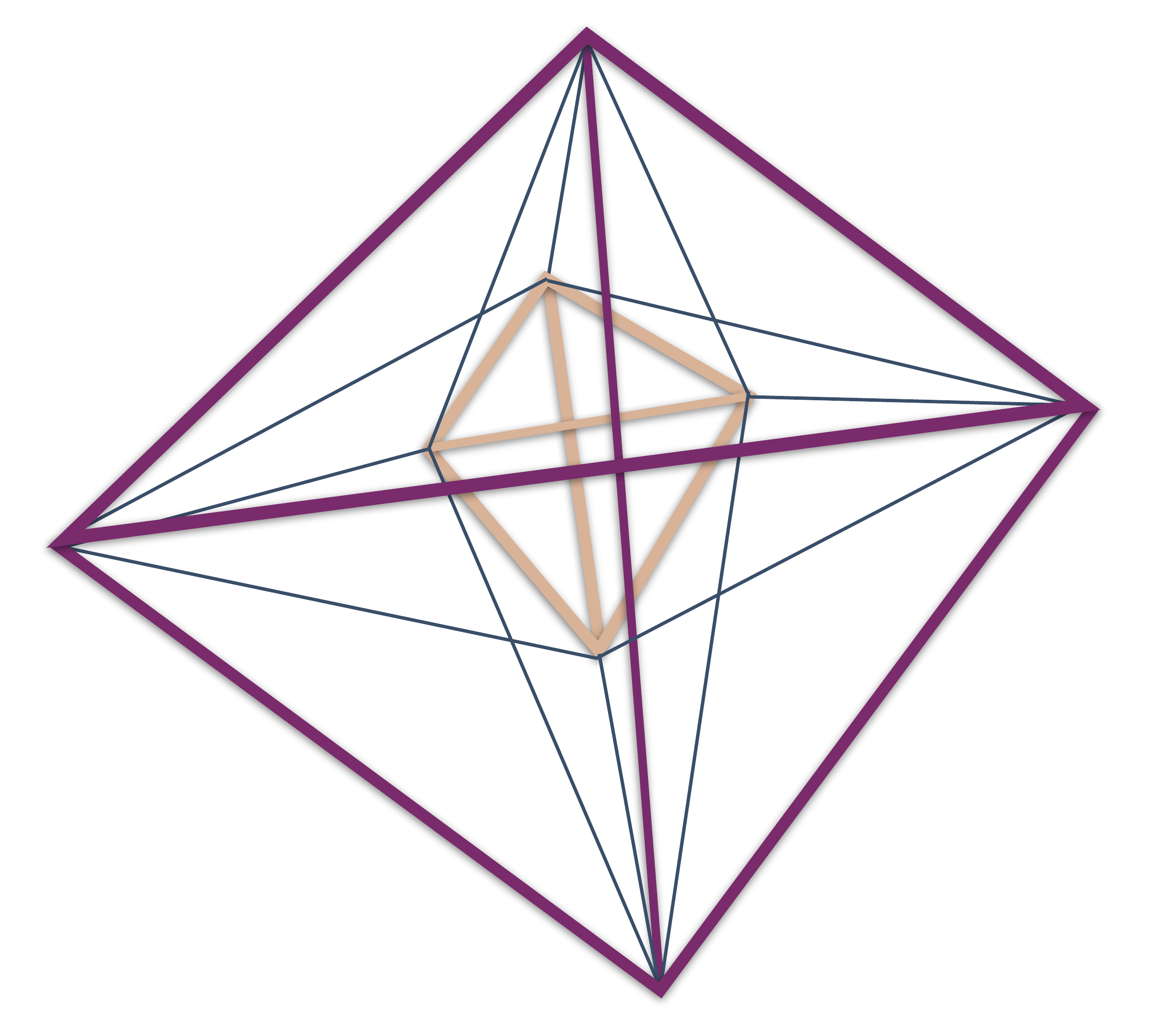}
    \caption{The triangulation of $\Sigma^t$. The brown tetrahedron $t_-$ is inscribed into the larger violet tetrahedron $t_+$. The blue segments connect vertices of the two tetrahedra radially.}
    \label{fig:triangulation-sigma11}
\end{figure}

\begin{figure}
\centering
  \subfigure[][]{%
  \label{fig:triangulation_sigma_t_plain}%
  \includegraphics[scale=0.15]{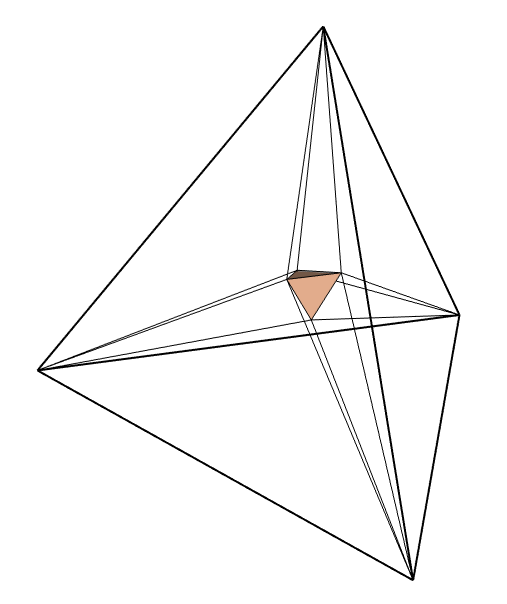}
  }%
  \hspace{22pt}
  \subfigure[][]{%
  \label{fig:triangulation_sigma_t_T-}%
  \includegraphics[scale=0.17]{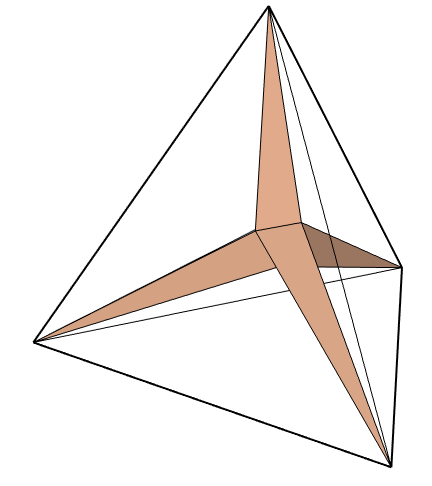}
  }%
  \hspace{29pt}
  \subfigure[][]{%
  \label{fig:triangulation_sigma_t_T+}%
  \includegraphics[scale=0.17]{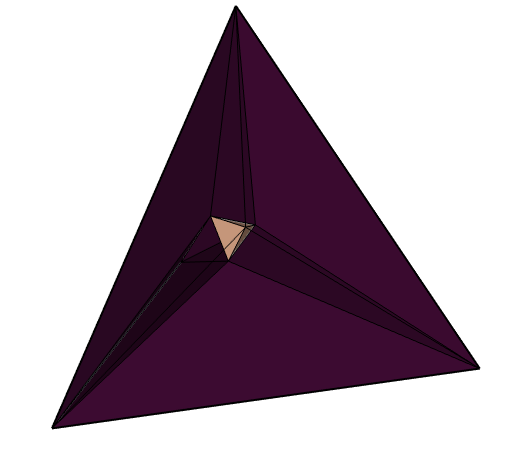}
  }
    \hspace{25pt}\subfigure[][]{%
  \label{fig:triangulation_sigma_t_Tab_alto}%
  \includegraphics[scale=0.15]{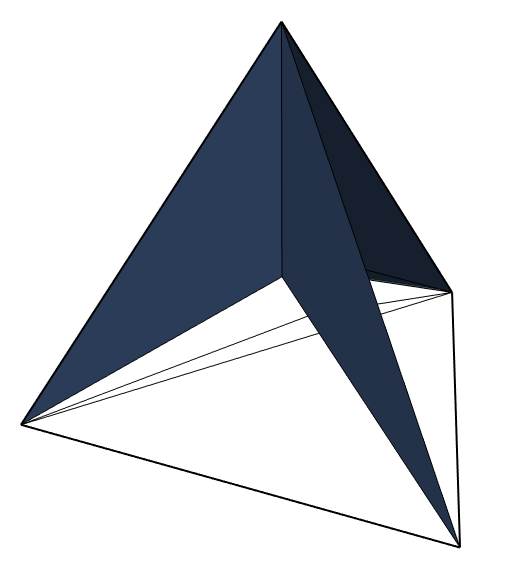}
  }%
  \hspace{25pt}
  \subfigure[][]{%
  \label{fig:triangulation_sigma_t_Tab_basso}%
  \includegraphics[scale=0.16]{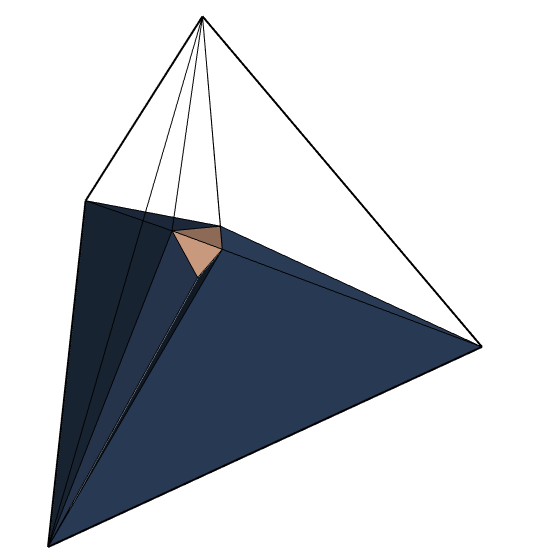}
  }%
  \hspace{19pt}
  \subfigure[][]{%
  \label{fig:triangulation_sigma_t_misto}%
   \includegraphics[scale=0.15]{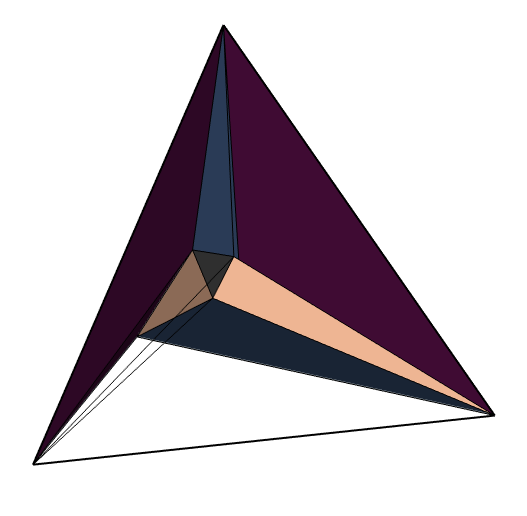}
   }\\
\caption{All images represent the triangulation of $\Sigma^{t}$, but with different tetrahedra highlighted. In \subref{fig:triangulation_sigma_t_plain} no tetrahedron is highlighted (the tetrahedron $t_-$ is in brown to remind that it is hollowed inside);
 in \subref{fig:triangulation_sigma_t_T-} the four $T^t_{-a}$ are highlighted;
 in \subref{fig:triangulation_sigma_t_T+} three $T^t_{+a}$ out of four are highlighted; 
 in \subref{fig:triangulation_sigma_t_Tab_alto} three $T^t_{ab}$ out of six are highlighted;
 in \subref{fig:triangulation_sigma_t_Tab_basso} the remaining three $T^t_{ab}$ are highlighted;
 in \subref{fig:triangulation_sigma_t_misto} two $T^t_{-a}$, two $T^t_{+b}$ and two $T^t_{cd}$ are highlighted.
  }%
\label{fig:triangulation_sigma_t}%
\end{figure}

Thanks to the $Z_2$ time reversal symmetry, the triangulations describing $\Sigma^p$ and $\Sigma^f$ must be topologically equivalent. For this reason, the same construction can be applied to both. A convenient triangulation for $\Sigma^t$, illustrated in Figures~\ref{fig:triangulation-sigma11} and~\ref{fig:triangulation_sigma_t}, is the following one. The placement of the smaller tetrahedron $t_-$ inside the bigger tetrahedron $t_+$, which can be chosen arbitrarily, is taken to be as in Figures~\ref{fig:triangulation-sigma11} and~\ref{fig:triangulation_sigma_t}\subref{fig:triangulation_sigma_t_plain}, such that the vertex $v_{- a}$ lies on the segment linking the centroid of the tetrahedron $t_+$ (which coincides with the centroid of the tetrahedron $t_-$) and the centroid of the face $\ell_{+ a}$. Then, each vertex $v_{+ a}$ of $t_+$ is linked to the three vertices of the triangle $\ell_{-a}$ (Figures~\ref{fig:triangulation-sigma11} and~\ref{fig:triangulation_sigma_t}\subref{fig:triangulation_sigma_t_plain}), creating 14 tetrahedra in total. 

We call $T^t_{+ a}$ (violet in Figure~\ref{fig:triangulation_sigma_t}\subref{fig:triangulation_sigma_t_T-}) the tetrahedron having $\ell_{+ a}$ as one of its faces and $T^t_{- a}$ (brown in Figure~\ref{fig:triangulation_sigma_t}\subref{fig:triangulation_sigma_t_T+}) the tetrahedron having $\ell_{- a}$ as one of its faces. Each of the six remaining tetrahedra (blue in Figures~\ref{fig:triangulation_sigma_t}\subref{fig:triangulation_sigma_t_Tab_alto} and~\ref{fig:triangulation_sigma_t}\subref{fig:triangulation_sigma_t_Tab_basso}) is bounded by two of the $T^t_{+a}$ tetrahedra and two of the $T^t_{-a}$ tetrahedra. Noting that the labels given to the $T^t_{+a}$ and $T^t_{-a}$ tetrahedra are such that each of the six remaining tetrahedra is bounded by a set of tetrahedra $T^t_{+b}$, $T^t_{+c}$, $T^t_{-d}$ and $T^t_{-e}$, with $b\neq c \neq d \neq e$, we can then label the six remaining tetrahedra as $T^{t}_{bc} \equiv T^{t}_{cb}$, where the labels $b$ and $c$ refer to $T^t_{+b}$ and $T^t_{+c}$. Clearly, from $T^{t}_{bc}$ one can readily trace back the other two tetrahedra $T^t_{-d}$ and $T^t_{-e}$. 

The full triangulation of $\Sigma$ is constructed identifying each $\ell_{\pm a}$ face of $\Sigma^p$ with the $\ell_{\pm a}$ face of $\Sigma^f$. This completely defines the triangulation of $\Sigma$.

The complication of the triangulation chosen is due to the non trivial topology of $\Sigma$ and from the computational opportunity of choosing a triangulation that respects the symmetries of the problem.


\subsection{The dual of the triangulation}

\noindent
In covariant LQG one works with the dual of a cellular decomposition of a spacetime region. More precisely, the spinfoam that captures the discretised degrees of freedom of the geometry is supported by the 2-skeleton of the dual of the cellular decomposition. The boundary of the spinfoam is the boundary spin-network, which is dual to the boundary triangulation. The graph $\Gamma_\Sigma$ of the spin-network is the two-skeleton of the dual of the boundary triangulation.

\begin{figure}[b]
\includegraphics[scale=0.056]{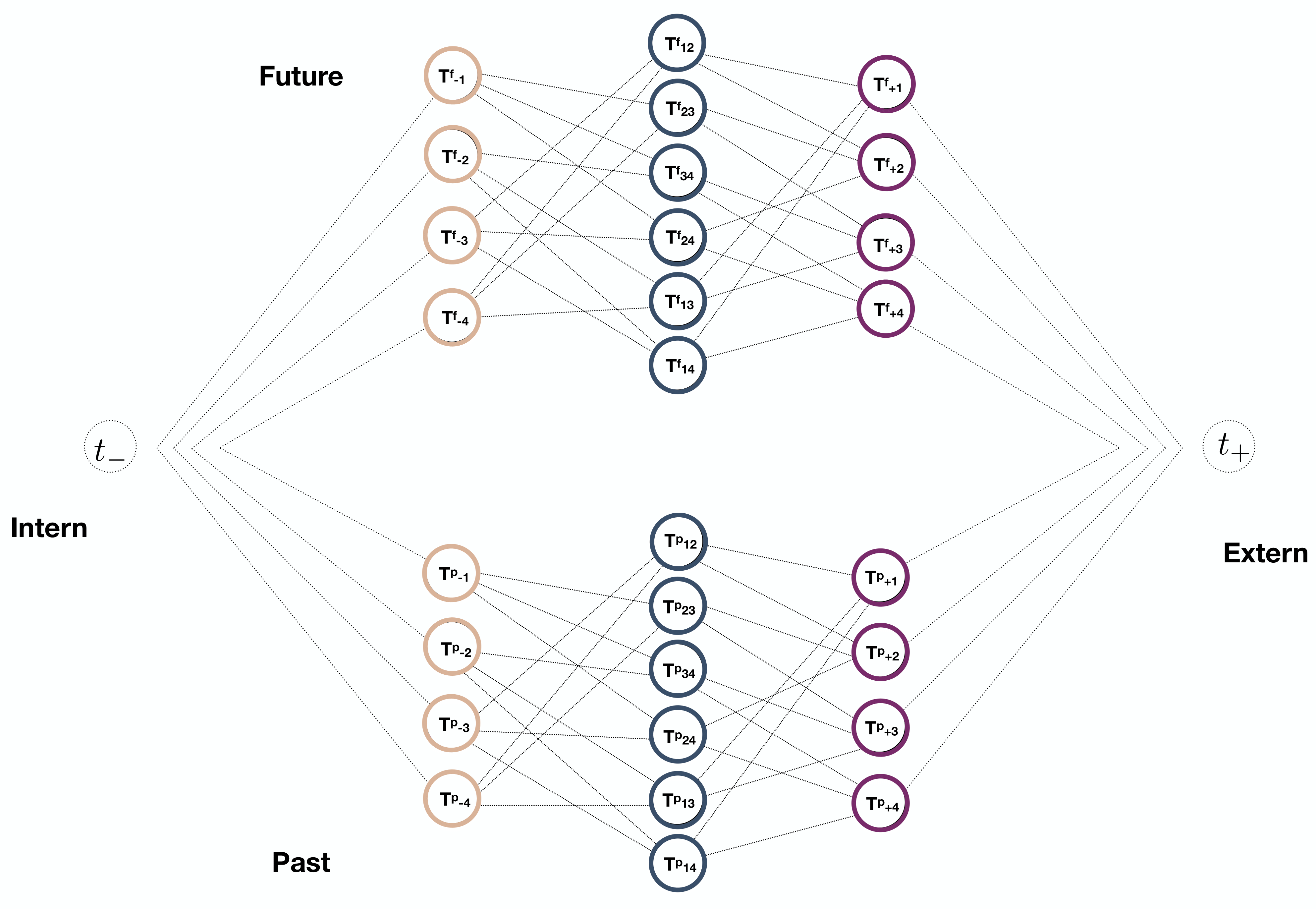}
\caption{Two-dimensional graph $\Gamma_\Sigma$ of the spin-network of $\Sigma$. The circles are nodes (dual to tetrahedra) and the segments are links (dual to the triangles).}
\label{fig:due_}
\end{figure}

The spin-network graph $\Gamma_\Sigma$ for the triangulation we have constructed, is illustrated in Figure~\ref{fig:due_}. Each circle is a node of the spin-network, and represents a tetrahedron, and each link joining two nodes represent a triangle separating two tetrahedra. (Intersections of links in this two-dimensional graph representation have no meaning.)

\begin{figure}[b]

  \includegraphics[keepaspectratio=true,scale=0.17]{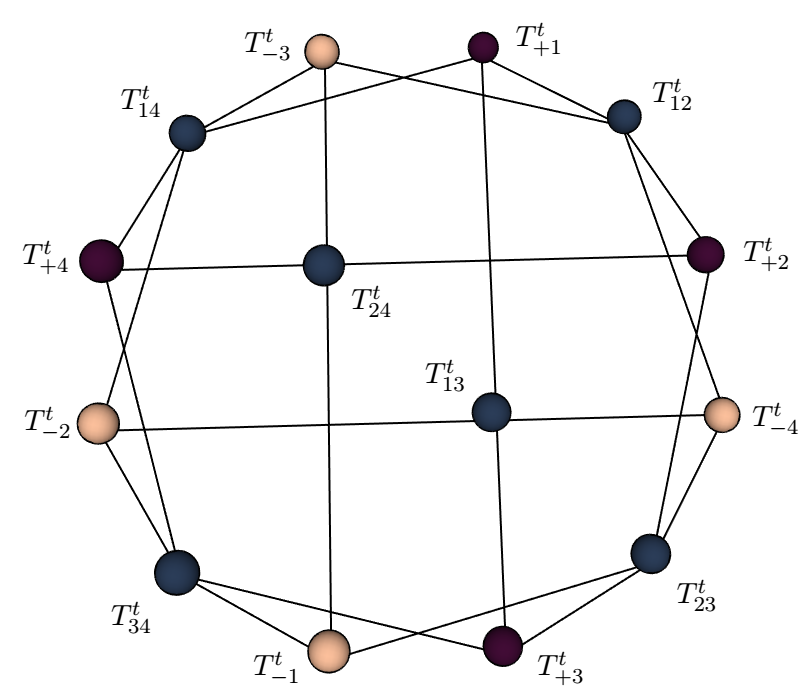}
  \hspace{1cm}
\raisebox{4mm}{\includegraphics[keepaspectratio=true,scale=0.17]{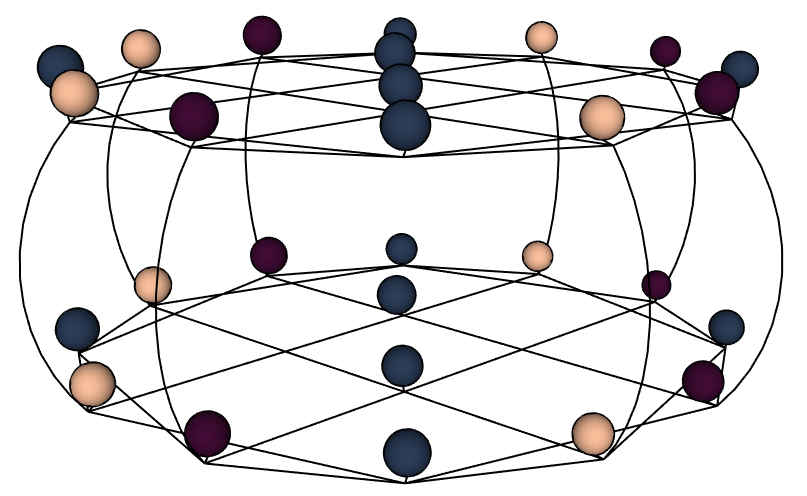}}

\caption{The left figure portrays a two-dimensional representation of the dual of the triangulation of $\Sigma^t$ and the right figure portrays a three-dimensional representation of the dual $\Gamma_\Sigma$ of the full triangulation of $\Sigma$, with labels omitted; one can easily label the right figure reading the different labels from the left figure and using $t=p$ for the bottom and $t=f$ for the top.}
\label{fig:dual_sigma_t_2D}     


\end{figure}

Since the information carried by the graph of a spin-network is only in its topology, as long as the latter remains unchanged, the graph can be deformed at will. Although the graphical representation of the dual graph $\Gamma_\Sigma$ in Figure~\ref{fig:due_} is completely fine to represent the topological information of the spin-network, it is not the best choice to manifestly represent all of its symmetries. A more symmetrical representation is the one in Figure~\ref{fig:dual_sigma_t_2D}.

Although the graph $\Gamma_\Sigma$ is quite complicated, thanks to the symmetries of the problem it has only two kinds of nodes that are topologically distinct: the $T^t_{\pm a}$ nodes and the $T^t_{ab}$ nodes. The symmetries act by permuting the $a$ indices and exchanging $p$ with $f$ or $+$ with $-$. Geometrically, the $T^t_{+ a}$ nodes differ from the $T^t_{- a}$ ones, as the last symmetry is not geometrical. For the same reasons, there are only four kind of links up to geometrical symmetries (two kind up to topological symmetries). These correspond to:

\begin{itemize}

\item the 4 links $\ell_{+ a}$ dual to triangles forming the discretised sphere $S_+$;

\item the 4 links $\ell_{- a}$ dual to triangles forming the discretised sphere $S_-$; together with the $\ell_{+ a}$ links they connect $\Sigma^p$ with $\Sigma^f$ (they are the vertical links in the second panel of Figure~\ref{fig:dual_sigma_t_2D});

\item the 24 links $\ell^t_{(+ a)(bc)}$ (12 for each $t$), with $b=a, \: c\neq a$, dual to the internal triangles separating the boundary tetrahedra $T^t_{+ a}$ (violet) from the internal tetrahedra $T^t_{bc}$ (blue);

\item the 24 links $\ell^t_{(- a)(bc)}$ (12 for each $t$), with $a\neq b\neq c$, dual to the internal triangles separating the boundary tetrahedra $T^t_{- a}$ (brown) from the internal tetrahedra $T^t_{bc}$ (blue).

\end{itemize}

The geometrical data that characterise the discretised geometry (and define coherent spin-network states) are the areas of the triangles and the angles between tetrahedra at these triangles. Hence, the relevant boundary data for the calculation are:

\begin{itemize}

\item the 2 areas $a_\pm$ of the internal and the external sphere $S_\pm$, which determine the areas associated to the links $\ell_{\pm a}$;

\item the 2 areas $A_\pm$ of the triangles dual to the $\ell^t_{(\pm a)(bc)}$ links;

\item the 2 (thin) angles $k_\pm$ between $\Sigma^p_\pm$ and $\Sigma^f_\pm$ at the internal and external sphere, which determine directly the angles associated to the links $\ell_{\pm a}$;

\item the two (thick) angles $K_\pm$ that depend on the extrinsic curvature of $\Sigma_\pm$ and that are associated to the triangles dual to the $\ell^t_{(\pm a)(bc)}$ links; the angles in $\Sigma^p$ have the opposite sign of the angles in $\Sigma^f$.

\end{itemize}

The extrinsic coherent state $\psi_{a_\pm,k_\pm,A_\pm,K_\pm}$ on the graph $\Gamma_\Sigma$ defined by the geometrical data $(a_\pm,k_\pm,A_\pm,K_\pm)$ represents the incoming and outgoing quantum states that correspond to the external classical geometry~\cite{Rovelli2015}. The LQG transition amplitude between coherent states will be a function of eight real numbers, with rather clear geometrical interpretation:
\be
W(a_\pm,k_\pm,A_\pm,K_\pm)=W(\psi_{a_\pm,k_\pm,A_\pm,K_\pm})\, ,
\label{amplitude}
\ee
where $W(\psi)$ for an arbitrary state $\psi$ in the boundary quantum state is defined in~\cite{Rovelli2015}.

In turn, these eight numbers $c_n=(a_\pm,k_\pm,A_\pm,K_\pm)$ depend on the geometry of $\Sigma$ described in the previous section. Hence, they depend on the four parameters $m, r_\pm$, and $v$ defined above. This defines the amplitude for the black-to-white hole transition as a function of these parameters:
\be
W(m, r_\pm, v)=W(c_n(m, r_\pm, v))\, .
\label{amplitude_2}
\ee
Our last task is to compute the functions $c_n(m, r_\pm, v)$.

\subsection{Discrete geometrical data}

\noindent
There is no unique or right way to assign discrete geometrical data to the graph $\Gamma_{\Sigma}$ starting from the continuous geometry of $\Sigma$. Each choice defines a different approximation of the continuous geometry and it has its own strengths and its own weaknesses. In this section we will introduce a convenient set of discrete geometrical data approximating the continuous geometry of $\Sigma$. We will discuss the discrete geometry of the triangulation approximating $\Sigma^p$. The discrete geometry of the triangulation approximating $\Sigma^f$ is simply related to the first one by a time reversal transformation.

The area of the spheres $S_\pm$ is directly determined by the radii $r_\pm$. Since the four triangles $\ell_{\pm a}$ bounding the tetrahedra $t_\pm$ that discretise the spheres $S_\pm$ are equal by symmetry, we take their area $a_{\pm}$ to be one fourth of the area of the spheres, that is
\be
     a_{\pm} = \pi r_\pm^2. 
\ee
The side length $b_\pm$ of the tetrahedron $t_{ \pm}$ is then trivially given by
\be
 b_\pm =  \sqrt{\frac{4\pi}{\sqrt{3}}} \, r_\pm \,.
\label{eq:b+-_side_t+-}
\ee

The line element $\dd s^2$ on $\Sigma^p$ can be written as 
\be
\dd s^2= f^2 (r) \, \dd r^2+r^2\,\dd\Omega^2\,,
\label{equation:metric_f}
\ee
where
\[
f^2 (r)=\beta\left(2-\beta\left(1-\frac{2m}{r}\right)\right)
\]
on $\Sigma^p_+$ (see equation~(\ref{equation:metric+})) and $f^2 (r)=1$ on $\Sigma^p_-$ (see equation~(\ref{equation:metric-})). We approximate this line element as
\be
\dd s^2= \alpha^2 \, \dd r^2+r^2\,\dd\Omega^2\,,
\label{equation:metric_alpha}
\ee
where $\alpha$ is a constant that needs to be determined. In order to do so, let $V_{\Sigma^p}$ and $V^{\alpha}_{\Sigma^p}$ be the volume of the surface $\Sigma^p$ whose intrinsic geometry is given respectively by equation \eqref{equation:metric_f} and \eqref{equation:metric_alpha}. The volume $V_{\Sigma^p}$ is given by
\begin{align*}
        V_{\Sigma^p} &=  \int_{\Sigma^t} \dd^3x \sqrt{|\det g^{(3)}|} \\
      &=  \int_{\Sigma^t_+} \dd r\, \dd\theta\, \dd\phi\, r^2 |\sin \theta| \sqrt{\left|\beta\left(2-\beta\left(1-\frac{2m}{r}\right)\right)\right|}\\
      & \quad +  \int_{\Sigma^t_-} \dd r\, \dd\theta\, \dd\phi\, r^2 |\sin \theta| \\
      &= 4 \pi \int_{\Sigma^t_+} \dd r\,  r^2  \sqrt{\left|\beta\left(2-\beta\left(1-\frac{2m}{r}\right)\right)\right|}\\
      & \quad + \frac{4 \pi}{3} \left(r_{S^p}^3-r_-^3\right) \, .
\end{align*}
The integral over $\Sigma^p_+$ can be computed explicitly (with computer algebra) in the case in which $\beta$ is fixed by the continuity at $S^p$. We do not give the explicit expression here. 
The volume $V^{\alpha}_{\Sigma^p}$ is instead given by
\begin{align*}
      V^{\alpha}_{\Sigma^p} &=  \int_{\Sigma^t} \dd^3x \sqrt{|\det g_{\alpha}^{(3)}|} = \int_{\Sigma^t} 
      \dd r\, \dd\theta\, \dd\phi\, \alpha \, r^2 |\sin \theta| \\
      &=  \frac{4 \pi  \alpha}{3} \left(r_{+}^3 - r_-^3\right) \, .
\end{align*}
We fix the value of $\alpha$ by requiring that $V^{\alpha}_{\Sigma^p}=V_{\Sigma^p}$. This explicitly gives the value of $\alpha$ as a function of the spacetime free parameters $(m,r_+,r_-,\beta)$.

We can now assign discrete geometrical data to the triangulation starting from the continuous instrinsic geometry in equation~(\ref{equation:metric_alpha}). Let us first consider the $\alpha=1$ case. When $\alpha=1$ the line element describes flat space and the intrinsic discrete geometry of the triangulation is completely determined by the side lengths $b_\pm$. Let $\widetilde{d}^{\,t_\pm}_v$, $\widetilde{d}^{\,t_\pm}_f$ and $\widetilde{h}_{t_\pm}$ be respectively the distance between the centroid and a vertex of the tetrahedron $t_\pm$, the distance between the centroid and a face of the tetrahedron $t_\pm$ and the height of the tetrahedron $t_\pm$. Basic geometry shows that
\be 
\widetilde{d}^{\,t_\pm}_v = \sqrt{\dfrac{3}{8}} \, b_\pm\, ,\ \  
\widetilde{d}^{\,t_\pm}_f = \dfrac{1}{\sqrt{24}} \, b_\pm\, 
, \ \  
\widetilde{h}_{t_\pm} = \sqrt{\dfrac{2}{3}} \, b_\pm\, .
\ee
The height $\widetilde{h}_{T_\pm}$ of a tetrahedron $T^p_{\pm a}$ (they are all equal by symmetry) relative to the base $\ell_{\pm a}$ can be expressed as
\be
\widetilde{h}_{T_+} = 
\widetilde{d}^{\,t_+}_f - \widetilde{d}^{\,t_-}_v 
\, \quad \text{and} \quad 
\widetilde{h}_{T_-} =
\widetilde{d}^{\,t_+}_v - \widetilde{d}^{\,t_-}_f \, .
\ee 
It is then possible to define the volumes of all the tetrahedra, except the $T^p_{ab}$ ones, as
\be
\widetilde{V}_X=\dfrac{1}{3} \, \widetilde{h}_X\, a_X\, ,
\label{equation:volumes_flat}
\ee
where $X=\{ t_+,t_-,T_{+}, T_{-} \}$, $a_{t_+}=a_{T_+}=a_+$ and $a_{t_-}=a_{T_-}=a_-$. Finally, the volume $\widetilde{V}_{T}$ of a $T^p_{ab}$ tetrahedron (they are all equal by symmetry) is given by
\be
\widetilde{V}_T=\dfrac{1}{6} \left( \widetilde{V}_{t_+} - \widetilde{V}_{t_-} -4 \widetilde{V}_{T_+} -4 \widetilde{V}_{T_-}     \right)\, .
\label{equation:volume_Tab_flat}
\ee
This completely determines the intrinsic discrete geometry of the triangulation in the flat case in terms of $b_\pm$.

When $\alpha \neq 1$ the line element in equation~(\ref{equation:metric_alpha}) describes a three-dimensional cone. We are only interested in the region of the cone in which $r\in [ r_- , r_+ ] $. Hence, in this approximation, $\Sigma^p$ turns out to be locally flat. However, it cannot be embedded in a flat three-dimensional space in the same way in which a two-dimensional cone cannot be embedded in a flat two-dimensional space.

Let us focus on the three-dimensional curved geometry defined by the line element in equation~(\ref{equation:metric_alpha}). The consequence of the presence of $\alpha$ is simply a stretching of the radial lengths with respect to the geometry discussed in the $\alpha=1$ case while the tangential lengths remain fixed. In analogy with the $\alpha=1$ case we can then define the following quantities:
\be
d^{\,t_\pm}_v= \alpha\, \widetilde{d}^{\,t_\pm}_v =
\sqrt{\dfrac{3}{8}} \,\alpha \, b_\pm\, ,
\ee
\vspace{-.3cm}
\be
d^{\,t_\pm}_f = \alpha \,\widetilde{d}^{\,t_\pm}_f = 
\dfrac{1}{\sqrt{24}} \,\alpha\, b_\pm\, ,
\ee
\vspace{-.3cm}
\be
h_{t_\pm}  =\alpha\, \widetilde{h}_{t_\pm} = 
\sqrt{\dfrac{2}{3}} \, \alpha\,b_\pm\, ,
\ee 
\vspace{-.3cm}
\be 
h_{T_+}= \alpha \, \widetilde{h}_{T_+} = \alpha \,\Big(\widetilde{d}^{\,t_+}_f - \widetilde{d}^{\,t_-}_v  \Big)\, ,
\ee 
\vspace{-.3cm}
\be 
h_{T_-}= \alpha \,\widetilde{h}_{T_-} =
\alpha\,\Big( \widetilde{d}^{\,t_+}_v - \widetilde{d}^{\,t_-}_f \Big) \, .
\ee
The volumes of all the tetrahedra, except the $T^p_{ab}$ ones, can be then written as
\be
V_X=
\dfrac{1}{3} \, h_X\, a_X =
\dfrac{1}{3} \, \alpha \,\widetilde{h}_X\, a_X
= \alpha\,\widetilde{V}_X\, ,
\label{equation:volumes_curved}
\ee
where $X=\{ t_+,t_-,T_{+}, T_{-} \}$, $a_{t_+}=a_{T_+}=a_+$ and $a_{t_-}=a_{T_-}=a_-$. Furthermore, since the curved counterpart of equation~(\ref{equation:volume_Tab_flat}) must still be valid, we can write 
\be
\begin{split}
V_T & = \dfrac{1}{6} \left(V_{t_+} - V_{t_-} -4 V_{T_+} -4 V_{T_-} \right) \\
& =\dfrac{\alpha}{6} \left( \widetilde{V}_{t_+} - \widetilde{V}_{t_-} -4 \widetilde{V}_{T_+} -4 \widetilde{V}_{T_-}     \right) \\
& = \alpha\,\widetilde{V}_T\, .
\end{split}
\label{equation:volume_Tab_curved}
\ee
The intrinsic discrete geometry of the triangulation is thus completely determined in terms of $b_\pm$ and $\alpha$. 

We are interested in the value of the areas $A_\pm$ of the triangles dual to the $\ell^p_{(\pm a)(bc)}$ links. These values are given by
\be
    A_\pm = \frac{1}{2} b_\pm h_\pm \, ,
    \label{eq:area_A}
\ee
where $h_\pm$ is the height of the triangles dual to $\ell^p_{(\pm a)(bc)}$ relative to $b_\pm$. Basic geometry shows that
\be
    h_\pm = \sqrt{ h^2_{T_\pm} + \frac{1}{12}b^2_\pm}\, .
    \label{eq:height_h_for_area_A}
\ee
So 
\begin{equation}
    \label{eq:A(alpha,r)}
    A_\pm = \pi r_\pm^2 \sqrt{\frac{\alpha^2}{18} \left( 1 - 3 \frac{r_\mp}{r_\pm} \right)^2 + \frac{2}{3}}
\end{equation}
With $\alpha$ fixed by the value of the total volume of $\Sigma^p$, equation \eqref{eq:A(alpha,r)} explicitly gives the value of the areas $A_\pm$ in terms of the spacetime free parameters $(m,r_+,r_-,\beta)$.

\begin{figure}
    \centering
    \includegraphics[scale =0.07]{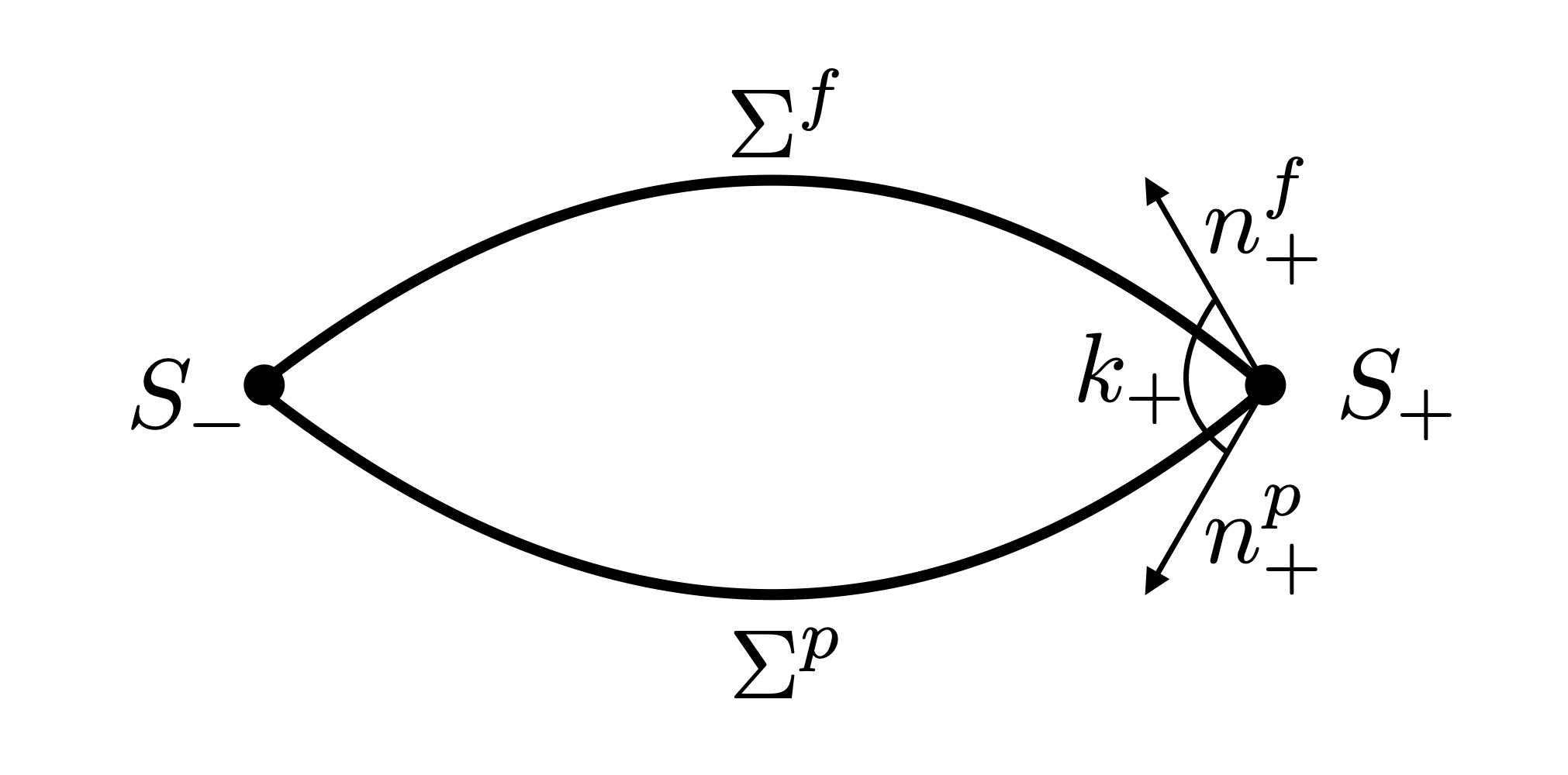}
    \caption{Definition of $k_+$}
    \label{fig:defk+}
\end{figure}

Let us now focus on the extrinsic discrete geometry. The angles $ k_\pm$, which are represented in Figure~\ref{fig:defk+}, are defined as
\be
    \cos k_\pm \overset{\rm def}= \left. \left(g^{\mu \nu} n^{\pm f}_\mu n^{\pm p}_\nu\right)\right|_{S_\pm}\, .
\ee
It is then straightforward to find
\be
    \cos k_+ = \frac{1+ \big[ (1-2m/r_+)\,\beta -1\big]^2}{|\beta(\beta-2-2m\beta/r_+)|(1-2m/r_+)}
\ee
and
\be
    \cos k_- =\frac{1+2m/r_-}{1-2m/r_-}\, .
\ee

The angles $K_\pm$ bear the extrinsic curvature of $\Sigma_\pm$. We choose to define them as the average of the extrinsic curvature, shared over the $12$ triangles $l^t_{(\pm a)(bc)}$: 
\be
    K_\pm = \frac{1}{12} \int_{\Sigma_\pm} k^a_a.
    \label{equation:K}
\ee
We have
\begin{multline}
        (k^+)^a_a =  \left(1 - \frac{2m}{r}\right) \frac{m\beta^{3/2} ( r(3-\beta) + 2m\beta)}{\sqrt{r^5(r(2-\beta)+2m\beta)}} \\
    - \frac{2}{r^2} \frac{r(1-\beta)+2m\beta}{\sqrt{\beta(2-(1-2m/r)\beta)}} 
\end{multline}
and 
\be
    (k^-)^a_a =  - \sqrt{\frac{m}{2r^3}}\left(3+\frac{2m}{r} \right).
\ee
For the time being, we leave the integral in equation~(\ref{equation:K}) unsolved.

Hence, we have found analytic expressions for the four areas $a_\pm$ and $A_\pm$ and the four angles $k_\pm$ and $K_\pm$ as functions of the four parameters $m, r_\pm$ and $\beta$ (the parameter $\beta$ can equivalently be traded for $r_{S^p}$ or $v$).

\section{Conclusions}

\noindent
The above construction defines the black-to-white hole transition amplitude $W(m, r_\pm, v)$ as a function of the physical parameters $(m, r_\pm, v)$ that characterise the transition and in terms of covariant LQG transition amplitudes. A number of questions, which we list here, remain open.
\begin{itemize}

\item To compute the amplitude $W(\psi_{a_\pm,k_\pm,A_\pm,K_\pm})$ to the first relevant order, we need to find a spinfoam bounded by $\Gamma_\Sigma$. This will be done in a forthcoming companion paper. 

\item The amplitude is then given by a complicated multiple group integral, which is hard to study. Asymptotic techniques, and in particular those recently developed in~\cite{Dona2020} are likely to be essential for this. Alternatively, a numerical approach, following~\cite{Dona2018,Dona2019} may provide insights in the amplitude. 
\item The question of eventual infrared divergences in the amplitudes and, eventually, how to deal with them, needs to be addressed.

\item To compute probabilities from amplitudes we have to address the problem of the normalisation. This can be solved using the techniques developed in the general boundary formulation of quantum gravity. See in particular~\cite{Oeckl2018}. Obviously the probabilities computed give the relative likelihood of a transition within the space of the parameter considered, and not the relative probability with respect to alternative scenarios on the end of the life of a black hole. 

\item We have taken a number of approximations which we do not control. Physical intuition suggests that the approximation given by disregarding a direct effect of Hawking radiation in the last phases of the evaporation, besides having already shrunk the horizon, may be of particular interest to check.

\item The black-to-white hole transition may have important astrophysical and cosmological implications. White hole produced by the transition of Planck size holes may be stable~\cite{Rovelli2018f} and form a component of dark matter. Alternatively, if the transition can happen at larger black hole masses, it may be related to cosmic rays and fast radio bursts~\cite{Barrau2014b,Barrau2016c,Barrau2017}. A control on the amplitude of this transition should help to shed light on these possibilities. 

\end{itemize}

\vspace{2mm}
\centerline{***}
\vspace{2mm}

\acknowledgments{\noindent We thank Alejandro Perez for useful comments. This work was made possible through the support of the  FQXi  Grant  FQXi-RFP-1818 and of the ID\# 61466 grant from the John Templeton Foundation, as part of the The Quantum Information Structure of Spacetime (QISS) Project (\href{qiss.fr}{qiss.fr}). }

\bibliographystyle{utcaps}

\bibliography{library.bib}

\end{document}

%% file: config.tex
\pdfoutput=1

\usepackage[autostyle, english = british]{csquotes}
\MakeOuterQuote{"}

\usepackage[english]{babel}

\usepackage{graphicx}
\usepackage{adjustbox}
\usepackage{subfigure}
\usepackage{capt-of}
\usepackage{amsmath,amssymb,amsfonts,tensor}
\usepackage{xcolor}

\usepackage{physics} 
\usepackage[percent]{overpic} 

\usepackage[colorlinks,citecolor=blue,linkcolor=blue,urlcolor=blue,pdfencoding=auto, psdextra]{hyperref}

\newcommand{\be}{\begin{equation}}
\newcommand{\ee}{\end{equation}}